# Interest of geophysical methods to determine the evolution and the spatial distribution of sedimentary deposits upstream of run-of-the-river dams (Upper Rhône, France)

Intérêt des méthodes géophysiques pour déterminer l'évolution et la distribution spatiale des dépôts sédimentaires en amont des barrages au fil de l'eau (Haut-Rhône, France)

Weit, A.[1], Winiarski, T.[1], Mourier, B.[1], Fretaud, T.[2] et Peteuil, C.[2]

[1] Univ Lyon, Université Claude Bernard Lyon 1, CNRS, ENTPE, UMR5023 LEHNA, F-69518, Vaulx-en-Velin, France,

[2] Centre d'Analyse Comportementale des Ouvrages Hydrauliques (CACOH) de CNR, 4 Rue de Chalon-sur-Saône, 69007 Lyon,

## RÉSUMÉ

Les ouvrages hydrauliques de type barrage de retenue influencent de manière directe ou indirecte le fonctionnement hydro-sédimentaire des rivières. Ils peuvent impacter la continuité sédimentaire du fleuve et peuvent créer un déséquilibre sédimentaire avec des zones d'accumulation de sédiments en amont et un manque de sédiments en aval de l'ouvrage. Les dépôts en amont constituent un enjeu fort pour les gestionnaires car ils peuvent impacter l'exploitation et/ou affecter la sûreté de l'ouvrage et provoquer un risque de sur-inondation. Les exploitants cherchent alors à déterminer les caractéristiques physiques de ces dépôts pour déterminer leur potentiel de remobilisation. Habituellement, ces dépôts sédimentaires sont décrits à l'aide de données bathymétriques, d'image sonar ou de prélèvements (dragage, carottage). Pour cette étude de nouvelles mesures issues de deux méthodes géophysiques (acoustique et électromagnétique) ont été mises en œuvre en complément des suivis courants. L'apport des techniques géophysiques est indéniable pour améliorer la caractérisation de ces dépôts. Ces mesures permettent de : i) définir les structures internes des sédiments, ii) déterminer la distribution spatiale de ces structures sur l'ensemble de l'aménagement iii) évaluer les volumes déposés lorsque l'on possède peu de données bathymétriques sur la retenue. Le couplage de ces méthodes permet ainsi de reconstruire l'évolution et les événements structurants des dépôts sédimentaires. Cette approche apporte des éléments complémentaires et importants pour les gestionnaires en termes d'exploitation et d'optimisation des scénarios de gestion des dépôts sédimentaires.

## ABSTRACT

Hydraulic structures such as dams have a direct or indirect influence on the hydro-sedimentary functioning of rivers. They can impact the sediment continuity of the river and can create a sediment imbalance with zones of sediment accumulation upstream and a lack of sediment downstream from the dam. Upstream deposits are a major issue for managers as they can impact the operation and/or affect the safety of the structure and induce extra-flood hazards. Operators therefore seek to determine the physical characteristics of these deposits to determine their remobilization potential. Usually, these sediment deposits are described using bathymetric data, sonar images or samples (dredging, coring). For this study, new measurements from two geophysical methods (acoustic and electromagnetic) were used in addition to these current monitoring methods. The contribution of geophysical techniques is undeniable to improve the characterization of these deposits. These measurements make it possible to: i) define the internal structures of the sediments, ii) determine the spatial distribution of these structures over the entire development, iii) evaluate the volumes deposited when little bathymetric data is available on the reservoir. The coupling of these methods thus makes it possible to reconstruct the evolution and the structuring events of the sedimentary deposits. This approach provides complementary and important elements for managers in terms of exploitation and optimization of sedimentary deposits management scenarios.

## KEYWORDS

Dam, geophysical methods, methodology, sediment deposits, Upper-Rhône,

# 1 INTRODUCTION

The accumulation of sediments upstream of dams is an important issue because it can cause a loss of water storage capacity or weaken the structures. This is the case for many reservoirs in the world but also in smaller reservoirs such as run-of the river dams or derivation dams. Knowledge of the sediment depositstructures and their evolution over time improve the efficient management. Furthermore, determining the physical characteristics of the sediment such as the grain size distribution in the reservoir, sedimentation processes, rates and volumes are important for its subsequent management.

The aim of this study is to investigate how and which geophysical methods can be integrated in a methodology to obtain a detailed view of the sedimentation processes upstream of dams or in shallow continental water bodies.

# 2 GEOPHYSICAL METHODS IN SHALLOW CONTINENTAL WATERS

The use of geophysical methods in continental waters is limited by shallow water depth and/or narrow width. To be applicable in these areas, it is essential to adapt the instrumentation on small boats and only methods suitable for shallow water can be considered. Given these difficulties, there are few studies that highlight the deployment of these methods in continental water bodies and in particular their potential in the context of dams, rivers and shallow lakes.

Geophysical methods such as echosounders or sonar instruments to obtain information about river bottom topography are already well integrated in the regular monitoring of rivers, lakes or reservoirs. Depending on the water level we propose the two further methods: the Sub Bottom Profiler (SBP) and the Ground Penetrating Radar (GPR), provide an image of sediment structures over several meters deep. A SBP is an acoustic instrument that operates at very low frequencies to penetrate the sediment and scan the underlying layers. Short sound pulses are sent downwards and are reflected by the different sediment structures. SBP measurements are preferentially used for the study of sedimentary deposits under large water columns but can be used in shallower environments with depths larger than >3m. In deep waters, SBP measurements can penetrate up to several tens of meters into the sediment. The presence of aquatic vegetation as well as the presence of gas linked to the degradation of organic matter in the sediment disrupt the penetration and reception of the signal. The second method is the Ground Penetrating Radar which is commonly used on terrestrial sediments but also applicable in shallow water levels and then called Water Ground Penetrating Radar (WGPR) based on electromagnetic radiation to image the subsurface.

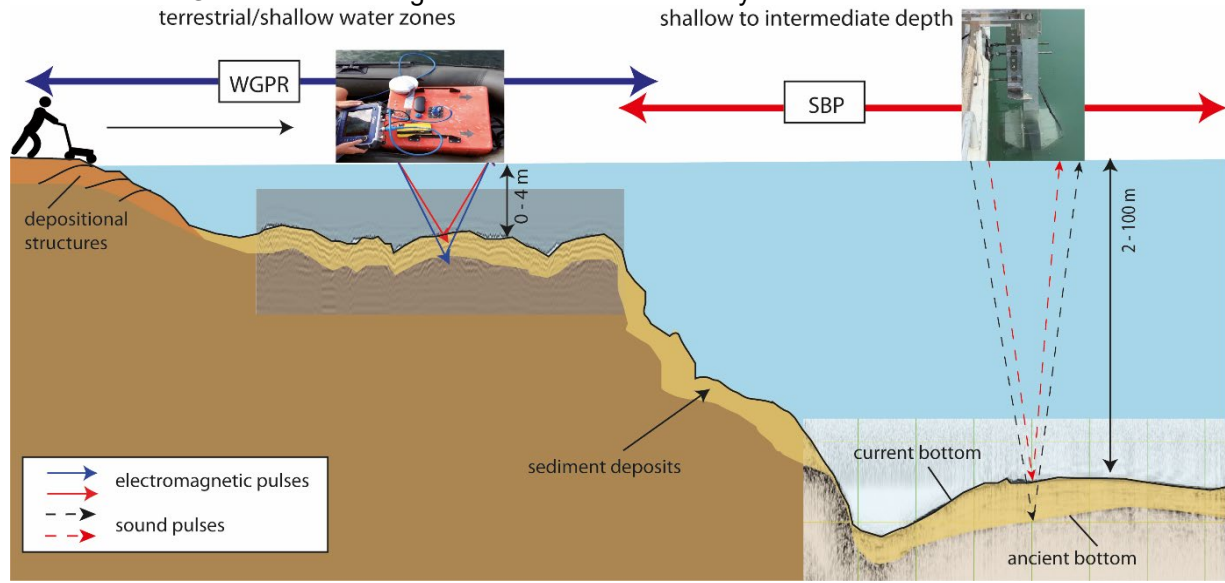

Figure 1. Application summary of the proposed geophysical methods SBP and GPR in shallow zones to investigate the internal structures of deposited sediments. The GPR can be used from land to shallow water zones up to 4m. while the SBP can be used in the deeper zones of reservoirs, lakes or rivers.

The GPR method is usually used in terrestrial environments, but it can be used in aquatic environments in the case of small water depths (up to 3 meters). It provides high-resolution representations of sedimentary structures in coarse to fine sediments but with difficulties in penetrating in strictly clayey sediments. To characterize sediments in lake or river environments, the combined use of SBP and GPR therefore makes it possible to obtain complementary sedimentary structure results [1].

We extended the methodological approach described by different authors [2,3]. The geophysical measurements (SBP, GPR) in combination with the data from sediment cores, bathymetric data and discharge make it possible to: i) define the internal structures of the sediments, ii) determine the spatial distribution of these structures over the entire reservoir, iii) evaluate the volumes deposited when little bathymetric data is available on the reservoir. This methodology allows identifying the zones at risk with respect to sediment accumulation. We tested this approach on several dam reservoirs in the Upper Rhône and were able to identify different sediment units and their spatial distribution within the reservoir.

## 3  RESULTS AND IMPLICATIONS

Combining the geophysical results with existing data on the different study sites on the Upper Rhône allows us to reconstruct the emplacement and evolution of the sediment deposits. An example of this method is shown in Figure 2 for the Chautagne reservoir (Upper-Rhône).

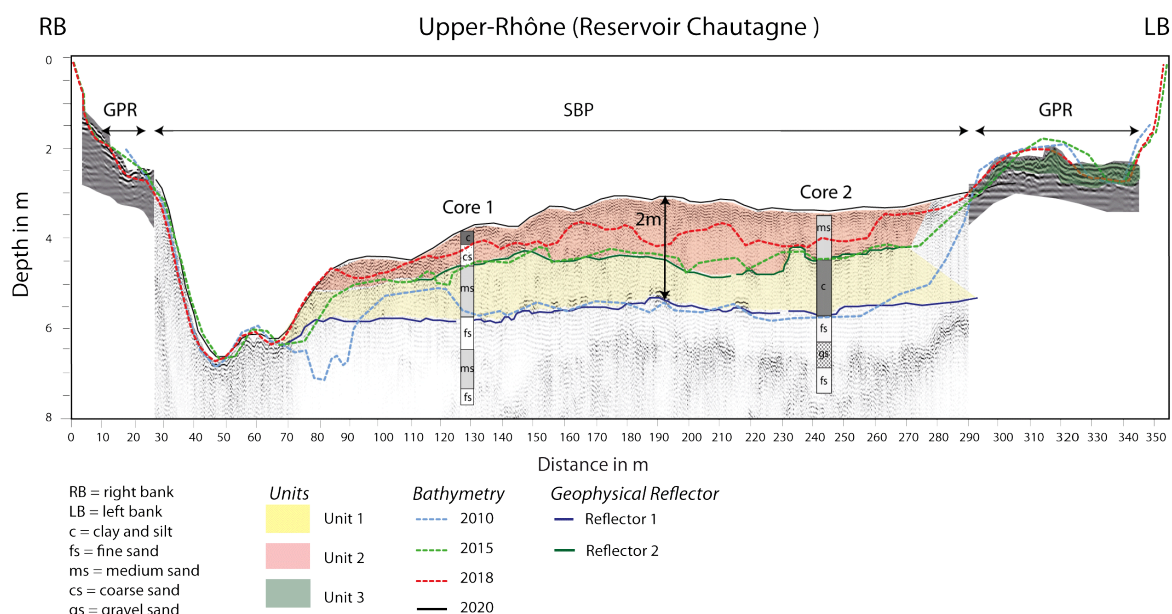

*Figure 2. Application of the geophysical methods on a transect located in the reservoir of Chautagne, Upper-Rhône.*

From the geophysical signal 3 different sediment units could be identified. Comparing these units to existing bathymetry data as well as sediment cores and hydraulic information we were able to determine two distinct events that marked the sedimentological evolution of the deposits having enough potency to create a distinct geophysical reflector. These events included a dredging event in 2009 where the sediments were compacted due to the weight of the dredging machine and the second event a flood period that created an erosion surface. The results from geophysical methods and the identification of the subsurface sediments can help to optimize field campaigns and specifically the position of new sediment cores and to lower management costs. This approach therefore provides complementary and important elements for managers in terms of exploitation and optimization of sedimentary deposit management.

## LIST OF REFERENCES